\documentclass[twocolumn,floats,showpacs,superscriptaddress,pre]{revtex4}
\usepackage{amssymb}
\usepackage{graphicx}
\usepackage{dcolumn}
\usepackage{amsmath}
\usepackage{bm}
\usepackage{epsfig}
\begin{document}
\title{Stochastic cellular automata model of neural networks}
\author{A. V. Goltsev}
\affiliation{Departamento de F{\'\i}sica da Universidade de Aveiro,
I3N, 3810-193 Aveiro, Portugal} \affiliation{A.F. Ioffe
Physico-Technical Institute, 194021 St. Petersburg, Russia}
\author{F. V. de Abreu}
\affiliation{Departamento de F{\'\i}sica da Universidade de Aveiro,
I3N, 3810-193 Aveiro, Portugal}
\author{S. N. Dorogovtsev}
\affiliation{Departamento de F{\'\i}sica da Universidade de Aveiro,
I3N, 3810-193 Aveiro, Portugal} \affiliation{A.F. Ioffe
Physico-Technical Institute, 194021 St. Petersburg, Russia}
\author{J. F. F. Mendes}
\affiliation{Departamento de F{\'\i}sica da Universidade de Aveiro,
I3N, 3810-193 Aveiro, Portugal}

\begin{abstract}
We propose a stochastic dynamical model of noisy neural networks
with complex architectures and discuss activation of neural networks
by a stimulus, pacemakers and spontaneous activity. This model has a
complex phase diagram with self-organized active neural states,
hybrid phase transitions, and a rich array of behavior. We show that
if spontaneous activity (noise) reaches a threshold level then
global neural oscillations emerge. Stochastic resonance is a
precursor of this dynamical phase transition. These oscillations are
an intrinsic property of even small groups of 50 neurons.
\end{abstract}


\maketitle
%%%%%%%\email{sdorogov@fis.ua.pt}
%%%%%%%\email{goltsev@fis.ua.pt}
%%%%%%%\email{jfmendes@fis.ua.pt}
%%\date{}

\section{Introduction}
\label{intro}

Understanding the dynamics and structure of neuronal networks is a
challenge for biologists, mathematicians and physicists. Neurons
form complex networks of connections, where dendrites and axons extend,
ramify, and form synaptic links between neurons. Due to long axons
the structure of a typical neuronal network has small-world
properties \cite{ws98, ab02,Sporns04,bghw04}. In particular,
neuronal networks in mammalian brains have short path lengths, high
clustering coefficients, degree correlations and skewed degree
distributions \cite{Sporns04}. Complex architectures of this kind
are known to strongly influence processes taking place on networks
\cite{dgm08,blmch06,ad-gkmz:08}. Complex wiring of neurons may be
important for the emergence of oscillations and synchrony in the brain
\cite{bghw04}.
%%Now the
%%function and mechanism of neural oscillations are topic
%%\cite{bghw04,sp06}. Noise is considered to be a possible mechanism
%%of oscillations \cite{Ermentrout08,Faisal08}.
Apart from this highly heterogeneous and compact structure, neural
networks are noisy \cite{Ermentrout08}. This makes a stochastic
approach to neuronal activities unavoidable
\cite{Ermentrout08,Faisal08}. Intuitively, noise is damaging,
however in neural networks noise can play a positive role,
supporting oscillations and synchrony \cite{Ermentrout08,Faisal08}
or causing stochastic resonance \cite{moss04,ma09}. According to
experimental data, oscillations and stochastic resonance may be
considered as ``noise benefits'' \cite{ma09}. The origin of these
phenomena, mechanisms and functions of oscillations in neural
networks are topical problems of great importance for the
understanding of brain function \cite{Ermentrout08,ma09}.
%%Mechanisms and functions of oscillations in neural networks are topic problems extensively discussed in current investigations.
Cultured neural networks provide well-controlled systems for \emph{in
vitro} investigations \cite{Eckmann07}. Despite their simplicity,
these cultured networks demonstrate an extremely rich repertoire of
activity due to interactions between hundreds to millions of
neurons. However, at present there is no complete understanding of
the dynamics of even these very simple neuronal networks. Recent
investigations \cite{Eckmann07} reveal that global activation of
living neural networks induced by
%%electrical stimulation
a stimulus can be explained on the base of
%%. The researchers used a simple model on a directed random network based on
the concept of bootstrap percolation---a version of cellular
automata---without going into details of neuron dynamics.
%%Studies in Refs. \cite{Breskin06, Eckmann07, Soriano08} were focused
%%on the global activation of neural networks with neurons of onetype.

In the present paper we propose a stochastic cellular automata model
of noisy neural networks. Based on experimental data we assume that
activation processes are stochastic, i.e., neurons can be activated
with a certain probability either by an external stimulus,
spontaneously, or by fluctuating inputs from active presynaptic
neurons. These networks include two neural populations, excitatory
and inhibitory neurons, and have a complex network architecture,
i.e., the small world property and heterogeneity are taken into
account. We consider model neurons which fire regular trains of
spikes with a constant frequency. The stochastic dynamics of these
networks takes into account processes of spontaneous neural
activity, which plays the role of noise, the activation of neurons
by a stimulus or neural pacemakers, and interactions between
neurons. With this model we aim to understand the role of noise in
the emergence of oscillations and the origin of stochastic
resonance. Although the model is simple, it demonstrates various
patterns of self-organization of neural networks, hybrid phase
transitions, hysteresis phenomena, neural avalanches and a rich set
of dynamical phenomena driven by noise: decaying and stable
oscillations, and stochastic resonance.
%%We will demonstrate a close relationship between noise,
%%oscillations, synchronization, and stochastic resonance.
%%This model also shows a constructive role of noise which leads to
%%global oscillations which synchronize even small groups of neurons.
%%Our stochastic model of neural networks demonstrated a close
%%relationship between noise, oscillations, synchronization, and
%%stochastic resonance.

Using exact analytical methods and simulations of the stochastic
dynamics of this model, we demonstrate that noise can play a
constructive role in neural networks. We show that at a critical level of
noise a neural network undergoes a dynamical phase transition from a
state with incoherent neurons to a state with synchronized neurons
and global oscillations. Oscillations of neural populations emerge
if spontaneous neural activity (noise) is above a critical level.
Stochastic resonance is a precursor of global oscillations. At a
given spontaneous neural activity, a critical fraction of neural
pacemakers can also stimulate oscillations.
%%This support ideas discussed in \cite{Ermentrout08}.
We consider several mechanisms leading to global oscillations in
neural populations: the difference in dynamics of excitatory and
inhibitory neurons or the existence of synaptic delays. These
mechanisms lead to similar oscillations. We also show that global
oscillations are intrinsic properties of the neural networks under
consideration. One should note that these oscillations are nonlinear
waves with a certain amplitude and a specific shape which are
determined by the structural and dynamical parameters. They do not
depends on initial conditions in contrast to waves in linear
dynamical systems. We demonstrate that the network structure plays
an important role. In neural networks having the structure of
classical random networks the larger the connectivity the broader is
the region with global oscillations. Our simulations reveal that
oscillations are an intrinsic property of even small groups of
neurons. 50-1000 neurons display oscillations similar to infinitely
large networks despite stochastic fluctuations which are usually
strong in small networks. The proposed model also explains a
discontinuous transition in the activation processes of living
neural networks observed experimentally in \cite{Eckmann07}. Neural
avalanches precede this transition. Simulations support our
analytical solution.

\section{Model}
\label{model}
%%\emph{Model.}---

\begin{figure}[t]
%%[tbhd]
\begin{center}
\scalebox{0.33}{\includegraphics[angle=270]{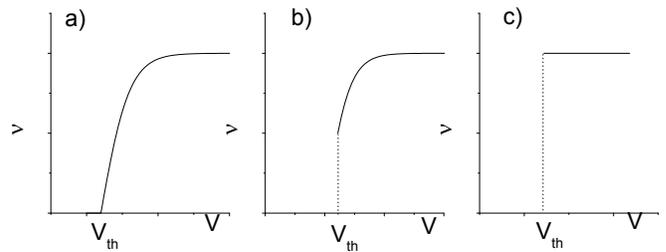}}
\end{center}
\caption{Firing rate $\nu$ versus input $V$: (a) type 1 neuron; (b)
type 2 neuron; (c) the step function approximation used in the
present paper.} \label{fig1}
\end{figure}

Neurons demonstrate various types of spiking behavior in response to
a stimulus at firing threshold, see, for example,
\cite{hodgkin:1948,izhikevich04,izhikevich06}. Type 1 neurons show a
continuous transition from an inactive state to an active state with
an arbitrary low firing rate when the input current is above a
threshold input (see Fig. \ref{fig1}a). For example, cortical
excitatory pyramidal neurons exhibit this behavior. Frequencies of
tonic spiking of type 1 neurons lie in the range from 2 Hz to 200
Hz, or can be even higher than 200 Hz. The maximum firing rate is
set by the refractory period of a neuron. Type 2 neurons show a
discontinuous transition to a nonzero firing rate above a threshold
input (see Fig. \ref{fig1}b). They fire in a relatively narrow
frequency band. For example, Hodgkin-Huxley neurons demonstrate type
2 neural excitability. Type 2 neurons fire spikes with frequency
about 40 Hz and higher. Fast-spiking inhibitory interneurons in the
rat somatosensory cortex fire
%%exhibit type 2 excitability with
in the frequency range 20-61 Hz \cite{thr04}. Neurons with type 2
dynamical behavior may play an important role in synchronization of
neural activity \cite{bpzb09}. Several models have been proposed
to describe the dynamics of individual neurons (see, for example,
\cite{izhikevich04,izhikevich06,r02,rtb04,lgns04}).
%%They give deterministic descriptions.

In the present paper we only consider regular
spiking neurons. We approximate the frequency-current response
by the step function (see Fig. \ref{fig1}c).
%%We assume that
Active excitatory and inhibitory neurons fire trains of spikes with
a constant frequency $\nu$ which is the same for all neurons and
does not depend on the input. If $\tau\nu > 1$ then during an
integration time $\tau$ (the membrane time constant) a postsynaptic
neuron receives $[\tau\nu]$ spikes from an active presynaptic
neuron, where [$A$] stands for the integer part of a number $A$. It
is assumed that the spike duration (about 1 ms) is much smaller than
$\tau$. The membrane time constant $\tau$ can range from 1 to 100 ms
\cite{Eckert-book97}. For example, for a typical integration time
$\tau = 10$ ms we must have $\nu >$100 Hz.
%%%%%%%%%%%%%%%%%%%%%%%%%%%%%%%%%%%%%%%%%%%%%%%%%%%%%%%%%%%%%%%%%%%%%
%%%%%%%%%%%%%%%%%%%%%%%%%%%%%%%%%%%%%%%%%%%%%%%%%%%%%%%%%%%%%%%%%%%%%%
%%%%%%%%%%%%%%%%%%%%%%%%%%%%%%%%%%%%%%%%%%%%%%%%%%%%%%%%%%%%%%%%%%%%%%
%%%%%%%%%%%%%%%%%%%%%%%%%%%%%%%%%%%%%%%%%%%%%%%%%%%%%%%%%%%%%%%%%%%%%%

Let us consider a neural network with two types of neurons:
excitatory and inhibitory neurons (see below). The total number of
neurons is $N$. The fractions of excitatory and inhibitory neurons
are $g_e$ and $g_{i}=1-g_e$, respectively. Neurons are linked by
directed edges and form a network with an adjacency matrix $a_{nm}$
where $n,m=1,2,... ,N$. An entry $a_{nm}$ is equal to 1 if there is
an edge directed from neuron $n$ to neuron $m$, otherwise
$a_{nm}=0$. Each neuron can be in either an active or inactive
state. Active neurons fire regular trains of spikes, as discussed
above. We assume that there is no phase correlation between trains
of spikes generated by different neurons. We define $s_{n}(t)=1$ if
neuron $n$ is active at moment $t$, and $s_{n}(t)=0$ if this neuron
is inactive. In our model, these binary variables play an auxiliary
role.
%% in contrast to other models with binary variables \cite{Amit,Derrida87,vs98,gl05}.
During the integration time $\tau$ a postsynaptic neuron receives
and integrates spikes from active presynaptic excitatory and
inhibitory neurons. First we consider the case $\tau\nu >1$. The
total input $V_{n}(t)$ (post-synaptic potential) at neuron $n$ is
the sum of inputs from nearest neighbor (presynaptic) neurons:
\begin{equation}
V_{m}(t)=[ \tau\nu ] \sum_{m}s_{m}(t)a_{mn}J_{mn},
 \label{input}
\end{equation}
where synaptic efficacy $J_{mn}=\pm J$ if neuron $m$ is excitatory
or inhibitory, respectively. We assume that all synapses of
excitatory neurons are excitatory, and all synapses of inhibitory
neurons are inhibitory. This is the so called Dale's principle
\cite{hi97}. Recently, the importance of Dale's principle for dynamics and pairwise correlations in neural networks was discussed by Kriener et al \cite{ktadr08}. In our model, the dynamics do not change
qualitatively if $|J_{mn}|$ are different for these two populations
of neurons. Note however that there are physiological reasons for
the fact that the magnitudes of inhibitory efficacies are usually
larger than excitatory efficacies (see, for example,
%%in Chapter 6 in
\cite{Eckert-book97}). Active excitatory (inhibitory) presynaptic
neurons give positive (negative) inputs to a postsynaptic neuron,
while inactive neurons give no input. For example, the input from
$k$ active excitatory and $l$ inhibitory neurons is $V=[\tau\nu]J k
- [\tau\nu] Jl$. We suppose that this input activates the
postsynaptic neuron if $V$ is at least a threshold value $V_{th}$.
This gives the following condition:
\begin{equation}
k-l \geq \Omega\equiv V_{th}/[\tau\nu]J
 \label{omiga}
\end{equation}
which we will use below. Notice that $\Omega$ is a dimensionless
parameter. The dimensionless threshold $\Omega$ is of the order of
15-30 in living neural networks \cite{Eckmann07} and about $30-400$
in the brain. Even if biological neurons have a variable threshold
\cite{izhikevich04}, for simplicity we assume that the threshold
does not depend on the prior activity.

In our stochastic model we assume that the states of neurons at
each moment $t$ are determined by the following rules:
\begin{itemize}
\item[(i)]
An excitatory (inhibitory) neuron is activated at a rate $f_e$
($f_i$) either by a stimulus or spontaneously (spontaneous activity).
%%In the former case, we assume that $f_e$ and $f_i$ depend on an applied
%%(typically, electric) field.
\item[(ii)]
In addition, an excitatory (inhibitory) neuron is activated at a
rate $\mu_{1e}$ ($\mu_{1i}$) by nearest neighbor
active neurons if the total input $V(t)$ at this neuron
%%excitatory  neurons (positive inputs) and inhibitory neurons (negative inputs)
is at least a threshold
value $V_{th}$, i.e., $V(t) \geq V_{th}$.
%%(For simplicity, we assume that this threshold is the same for all neurons.)
\item[(iii)]
An activated excitatory (inhibitory) neuron is inactivated (i.e., it
stops firing) at a rate $\mu_{1e}$ ($\mu_{1i}$) if the total input
$V(t)$ becomes smaller than $V_{th}$.
%%, i.e., $V(t) < \Omega$.
\item[(iv)]
An activated excitatory (inhibitory) neuron spontaneously stops
firing at rate $\mu_{2e}$ ($\mu_{2i}$).
\end{itemize}
%%Rules (ii) and (iii) introduce the
%%interactions between neurons.
%%These stochastic rules
%%of activation an deactivation of individual neurons
%%and sparse directed network structure differetiate our model from
%%attractor neural networks in \cite{Amit}.

In the brain, neurons receive
%%highly
fluctuating inputs and generate
%%irregular
spike trains \cite{Ermentrout08}.
%%This activity is spontaneous.
%%In our model,
We represent the activation
%%of a neuron by these
by fluctuating inputs as the stochastic process (ii) with the rates
$\mu_{1e}$ and $\mu_{1i}$ which can be of the order of the average
firing rate. This determines the time scale in the model. Even if
the total input is on average larger than $V_{th}$, it sometimes
falls below $V_{th}$. As a result, the neuron stops firing. Process
(iv) is meant to represent this. The biophysical meaning of the
model parameters, assumptions and approximations which are the basis
of our model, are discussed in Sec.~\ref{discussion}. For
other models with binary variables see
\cite{Amit,Derrida87,vs98,gl05} and in the review \cite{lgns04}.

In order to describe the dynamics of neural networks, we introduce a
probability $\rho_{n}^{(a)}(t)$ that neuron $n$ of type $a$ is
active at time $t$. Let us define the mean values of
$\rho_{n}^{(a)}(t)$ for excitatory, $a=e$, and inhibitory, $a=i$,
populations:
\begin{equation}
\rho_{a}(t)\equiv \sum_{n}\rho_{n}^{(a)}(t)/(g_{a}N),
 \label{act}
 \end{equation}
where the sum is over neurons of type $a$, $g_{a}$ is their
fraction. We name $\rho_{e}(t)$ and $\rho_{i}(t)$ ``activities'' of
the excitatory and inhibitory populations. On the other hand,
$\rho_{e}(t)$ and $\rho_{i}(t)$ are the respective probabilities that a
randomly chosen excitatory or inhibitory neuron is
active at time $t$. We consider neural networks whose structure is of a sparse
random uncorrelated directed network. These networks are small
worlds and can have an arbitrary degree distribution. They are often
considered as a good approximation to real networks \cite{ab02}. The
advantage of these networks is that they can be studied analytically
by use of mean-field theory and easily modeled for simulations.
However, they do not take into account the high clustering coefficient
and degree correlations of real neural networks \cite{Sporns04}.
Though the mean-field approach is based on the tree-like
approximation, it takes into account exactly the heterogeneity of
networks and large feedback loops \cite{dgm08}.
%%They are also easily modeled for simulations.

\section{Basic rate equations}
\label{basic}

Let us derive dynamical equations for the activities $\rho_{e}(t)$
and $\rho_{i}(t)$. We introduce the probabilities
$\Psi_{e}(\rho_{e}(t),\rho_{i}(t))$ and
$\Psi_{i}(\rho_{e}(t),\rho_{i}(t))$ that at time $t$ the total input
to a randomly chosen excitatory or inhibitory neuron, respectively,
is at least $\Omega$. If at time $t$ an excitatory neuron is
inactive, which takes place with probability $1-\rho_{e}(t)$, then
an external field activates this neuron at a rate $f_e$. This gives
a contribution
\begin{equation}
f_{e}[1-\rho_{e}(t)] \label{t1}
\end{equation}
to the rate $\dot{\rho}_{e}(t)\equiv d \rho_{e}(t)/dt$. If at time
$t$ the total input to an inactive neuron is at least $\Omega$,
which takes place with probability
$\Psi_{e}(\rho_{e}(t),\rho_{i}(t))$, then this neuron is activated
at the rate $\mu_{1e}$. This gives one more positive contribution
%%%
\begin{equation}
\mu_{1e}[1-\rho_{e}(t)] \Psi_{e}(\rho_{e}(t),\rho_{i}(t)).
\label{t2}
\end{equation}
If at time $t$ an excitatory neuron is active, which takes place
with probability $\rho_{e}(t)$, and the total input from activated
nearest neighbor excitatory neurons is smaller than $\Omega$, which
takes place with probability $1-\Psi_{e}(\rho_{e}(t),\rho_{i}(t))$,
then such an active neuron becomes inactive at the rate $\mu_{1e}$.
The active neurons also can stop spontaneously firing with rate
$\mu_{2e}$. These processes give two negative contributions:
\begin{equation}
-\mu_{1e} \rho_{e}(t) [1-
\Psi_{e}(\rho_{e}(t),\rho_{i}(t))]-\mu_{2e} \rho_{e}(t). \label{t3}
\end{equation}
Summing all contributions, we obtain %%leads to
a rate equation,
\begin{equation}
\dot{\rho}_{a}(t) = f_{a}-\nu_{a} \rho_{a}(t) +
\mu_{1a}\Psi_{a}(\rho_{e}(t),\rho_{i}(t)). \label{ei}
\end{equation}
Here $\nu_{a}\equiv f_{a}{+}\mu_{1a}{+}
\mu_{2a}$, and $a=e,i$.

To clarify the relative role of activation and deactivation
processes, we rewrite Eq.~(\ref{ei}) as follows: %%
\begin{equation}
\dot{\rho}_{a}/\nu_{a}\!\! =\!\!F_{a}(1{-}Q_{a}){-}
\rho_{a}{+}(1{-}F_{a})(1{-}Q_{a})\Psi_{a}(\rho_{e},\rho_{i}),
\label{ei2}
\end{equation}
where $\rho_{a}=\rho_{a}(t)$. The dimensionless parameters $F_{a}\equiv
f_{a}/(f_{a} + \mu_{1a})$ and $Q_{a}\equiv \mu_{2a}/\nu_{a}$
determine the relative strength of stimulation and the spontaneous
deactivation of neurons. The rates $\nu_{e}$ and $\nu_{i}$ set the
time scale.

The probabilities $\Psi_{e}$ and $\Psi_{i}$ are determined by the
network structure. Below we will study a directed classical random
graph which is the simplest and representative model of sparse
uncorrelated complex networks \cite{ab02,dgm08}. These random graphs
share the properties of sparse uncorrelated random networks with a
finite second moment of the degree distribution. They are small
worlds and have a mean shortest distance which increases as the
logarithm of the number of vertices, in contrast to a three
dimensional system where a mean shortest distance increases as the
cube root of the size. Due to simplicity, classical random graphs
are often used to study dynamics of systems having a complex network
structure \cite{ab02,dgm08,blmch06,ad-gkmz:08}. In contrast to real
networks, sparse random uncorrelated networks and in particular
classical random graphs have zero clustering coefficient due to
their tree-like structure and negligible (in some cases, weak)
degree-degree correlations between neighboring nodes in the infinite
size limit. Understanding the strength of the clustering and degree
correlations on dynamics of systems with complex network
architecture is an open problem in the theory of complex networks
\cite{dgm08,blmch06,ad-gkmz:08}. Recent investigations of various
dynamical models on complex networks show that in many cases
networks with clustering demonstrate dynamics qualitatively similar
to tree-like networks. In many cases degree-degree correlations also
do not qualitatively change the dynamics. This challenging problem
is discussed in detail in the recent review \cite{dgm08}.
%%Recent achievements in the complex network theory inspire optimism that new
%%theoretical approaches for studying of the role of clustering and
%%degree-degree correlations in complex networks will soon be found
%%\cite{newman09}.

In a classical random graph a directed edge between each pair of $N$
neurons is present with a given probability $c/N$. The parameter $c$
is the mean input and output degrees.
%%It is easy to show that
The probability $B_{n}(c)$ that a neuron has $n$ input edges is given by the
binomial distribution:
\begin{equation}
B_{n}(c)= C^{N}_{n}(\frac{c}{N})^{n}(1-\frac{c}{N})^{N-n}
\label{binomial}
\end{equation}
where $C^{N}_{n}=N!/(N-n)!n!$ is the binomial coefficient. We will
study analytically large networks with $N \gg 1$. In the infinite
size limit, $N \rightarrow \infty$, the binomial distribution
$B_{n}(c)$ approaches the Poisson distribution $P_{n}(c)$,
\begin{equation}
P_{n}(c)= c^{n}e^{-c}/n!, \label{poisson}
\end{equation}
%%(the input degree distribution).
which is more convenient for calculations. The probability that a
randomly chosen neuron has $k$ active presynaptic excitatory and $l$
active presynaptic inhibitory neurons is
$P_{k}(g_{e}\rho_{e}c)P_{l}(g_{i}\rho_{i}c)$. Hence, in the case
$\tau\nu >1$, we get
\begin{eqnarray}
\!\!\!\!\!\!\Psi_{e}(\rho_{e},\rho_{i}){=}\Psi_{i}(\rho_{e},\rho_{i}){=}\sum_{k \geq \Omega}\,\,\sum_{l=0
}^{k-\Omega} P_{k}(g_{e}\rho_{e}c)P_{l}(g_{i}\rho_{i}c)   \nonumber
\\[5pt]
\!\!\!\!\!\!\!\!\!\!\!\!\!\!\!\!\!\!\!\!\!\!\!\!\!\!\!\!\!\!\!\!\!\!\!\!\!\!\!\!\!\!
\!\!\!\!\!\!\!\!\!\!\!\!\!\!\!\!\!\!\!\!\!\!\!\!\!\!\!\!\!\!\!\!
{=}e^{{-}g_{e}\rho_{e}c}\sum_{k\geq
\Omega}\frac{(g_{e}\rho_{e}c)^{k}}{k!(k{-}\Omega)!}\Gamma(k{-}\Omega{+}1,g_{i}\rho_{i}c),
\label{Psi-ei}
\end{eqnarray}
where $\Gamma(k,x)$ is the upper incomplete gamma function and $\Omega$
is defined by Eq.~(\ref{omiga}). Notice that in the case of
classical random graphs we have used the fact that there are no
correlations between the number of input and output edges.

In the case $\tau\nu < 1$, during the integration time $\tau$ a
postsynaptic neuron receives only one spike or none from an active
presynaptic neuron. If the phase of a train of spikes is uncertain
then all we can say is that during the time interval $\tau$ with
probability $\tau\nu$ a postsynaptic neuron receives a spike from an
active presynaptic neuron. In turn, the probability that there is no
spike is $1- \tau\nu$. Let us assume that there is no phase
correlation between regular spiking neurons. This is a common
assumption at low activity rates \cite{AmitBrunel997}. The
probability that during time $\tau$ a neuron receives $k$ spikes
from uncorrelated $n$ regular spiking presynaptic neurons is
%%equal to
 \begin{equation}
C^{n}_{k}(\tau\nu)^{k}(1-\tau\nu)^{n-k}. \label{nk}
\end{equation}
%%where $C^{n}_{k}=n!/(n-k)!k!$ is the binomial coefficient.
In the
case of a classical random graph, the probability that during the
integration time $\tau$ a randomly chosen neuron receives $k$ spikes
from active excitatory or inhibitory neurons is given by the Poisson
distribution:
\begin{equation}
\sum_{n = k}^{\infty} P_{n}(g_{a}\rho_{a}c)
C^{n}_{k}(\tau\nu)^{k}(1-\tau\nu)^{n-k}=P_{k}(g_{a}\rho_{a}\tau\nu
c),  \label{k-spikes}
\end{equation}
where $a=e,i$ for excitatory and inhibitory neurons, respectively.
$k$ spikes from excitatory and $l$ spikes from inhibitory neurons
activate a postsynaptic neuron if $V=Jk-Jl \geq V_{th}$. Using the
probability Eq.~(\ref{k-spikes}), one can show that the
function $\Psi_{a}(\rho_{e}(t),\rho_{i}(t))$ in Eq.~(\ref{ei}) is
given by Eq.~(\ref{Psi-ei}) if the mean degree $c$ is replaced by
$\tau\nu c$, and a threshold $\Omega=V_{th}/J$ is used. Therefore,
the effective mean input degree is decreased while the effective
threshold is increased in comparison to the case $\tau \nu >1$.
%%We conclude that in the
%%case $\tau\mu < 1$, a larger level of spontaneous activity is
%%necessary to stimulate global neural oscillations.
Note that if trains of spikes generated by presynaptic neurons are correlated,
then Eq.~(\ref{nk}) is invalid. Spikes acting in concert can
activate a postsynaptic neuron more effectively.

One can use another approach. The stochastic rules (i)-(iv) lead to
a rate equation for the activity $\rho_{n}^{(a)}(t)$ of single
neuron $n$ of type $a$ with $q_{n}=\sum_{m}a_{mn}$ presynaptic
neurons for a given adjacency matrix $a_{nm}$:
\begin{eqnarray}
&& \!\!\!\!\!\!\!\!\!\!\!\!\!\!\!\!\!\!\! \dot{\rho}_{n}^{(a)}(t) {=} f_{a}{-}\nu_{a} \rho_{n}^{(a)}(t)
\nonumber
\\[5pt]
&& + \mu_{1a}\!\!\!\sum_{\{s_{m}=0,1\}}\!\!\!\! \Theta(V_{n}{-}V_{th}(n)) \prod_{m}[ a_{mn}\rho_{m}(s_{m},t)], \label{s-ei}
\end{eqnarray}
where $V_{n}=[\tau\nu]\sum_{m}s_{m}a_{mn}J_{mn}$ is the input at
neuron $n$ from presynaptic neurons $m$ at $[\tau\nu]>1$,
$\Theta(x)$ is the Heaviside step function,
$\rho_{m}(s_{m}{=}0,t)=1-\rho_{m}^{(a)}(t)$ and
$\rho_{m}(s_{m}{=}1,t)=\rho_{m}^{(a)}(t)$ are the probabilities that
presynaptic neuron $m$ is inactive or active at time $t$,
respectively. The last term in Eq.~(\ref{s-ei}) is the probability
that the input at neuron $n$ is at least the local threshold
$V_{th}(n)$ at time $t$. These equations describe a neural network
with a given adjacency matrix $a_{nm}$, arbitrary synaptic
efficacies $J_{nm}$ and arbitrary local thresholds $V_{th}(n)$. In
the case of the classical random graph in the infinite size limit,
for the uniform case $|J_{nm}|=1$ and $V_{th}(n)=V_{th}$, the set of
$N$ coupled nonlinear rate equations (\ref{s-ei}) can be reduced to
two coupled equations for the averaged activities $\rho_{e}$ and
$\rho_{i}$.  Summing over $n$ in Eq.~(\ref{s-ei}) and averaging over
the network ensemble, we arrive at Eqs.~(\ref{ei}) and
(\ref{Psi-ei}). We believe that the mean-field equation (\ref{ei})
is exact for sparse uncorrelated directed networks in the limit
$N\rightarrow \infty$. Our simulations of the model on classical
random graphs support this. Similar rate equations were derived for
disease spreading and contact processes on complex networks
\cite{p-sv01, Catanzaro:cbp05}.

Neural networks can also be activated by pacemakers (neurons that
permanently fire). Let excitatory and inhibitory pacemakers be
chosen with given probabilities $F_{e}$ and $F_{i}$ from excitatory
and inhibitory neurons, respectively.
The stochastic dynamics of remaining neurons (activities
$\widetilde{\rho}_{e}(t)$ and $\widetilde{\rho}_{i}(t)$) are
governed by rules (ii)-(iv). In the same way as for Eq.~(\ref{ei2}),
we obtain
\begin{equation}
\dot{\rho}_{a}/\nu_{a} =F_{a}{-}
\rho_{a} {+}
(1{-}F_{a})(1{-}Q_{a})\Psi_{a}(\rho_{e},\rho_{i}), \label{pm}
\end{equation}
where we define $\rho_{a}{\equiv}F_{a}
{+}(1{-}F_{a})\widetilde{\rho}_{a}(t)$, the total activity of the
neural population $a$, $a=e,i$.
%%Here $\nu_{a}\equiv \mu_{1a}{+}\mu_{2a}$ and $Q_{a}=\mu_{2a}/\nu_{a}$.
Equations~(\ref{ei2}) and (\ref{pm}) differ only by the first term
on the right-hand side. Thus, activation
%%of neural networks
by a stimulus or randomly chosen pacemakers produce similar effects.
A similar equation at $Q_{a}=0$ was derived
using another approach in \cite{vs98}.
%%From the physical point of view this
%%difference is obvious. Neurons activated by field can stop fire
%%spontaneously while pacemakers are firing forever by definition.  At
%%$Q_{a}=0$, i.e., if the spontaneous deactivation is absent,
%%Eq.~(\ref{pm}) is reduced to Eq.~(\ref{ei2}).
%%Our simulations of neural networks on classical random directed
%%graphs show that

In our model one can also take into account synaptic delays.
Introduce time $T_{ab}$ for the transmission of a nerve signal from
a neuron of type $a$ to a nearest neighbor neuron of type $b$, where
$a,b=e,i$. Then, in Eq.~(\ref{ei2}), replace
$\Psi_{a}(\rho_{e}(t),\rho_{i}(t))$ by
$\Psi_{a}[\rho_{e}(t-T_{ea}),\rho_{i}(t-T_{ia})]$.
Various sources of delays in the nervous system
%%, such as distance-dependent delays owing to the finite propagation velocity,
and their role in dynamics of neural networks were recently discussed by Ermentrout and Ko \cite{ek09}.
%%%%%%%%%%%%%%%%%%%%%%%%%%%%%%%%%%%%%%%%%%%%%%%%%%%%%%%%%%%%%%%%%%%%%
%%%%%%%%%%%%%%%%%%%%%%%%%%%%%%%%%%%%%%%%%%%%%%%%%%%%%%%%%%%%%%%%%%%%%%
\begin{figure}[t]
\begin{center}
\scalebox{0.23}{\includegraphics[angle=0]{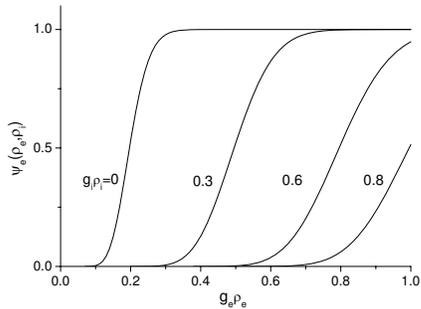}}
\end{center}
\caption{ Function $\Psi_{a}(\rho_{e},\rho_{i})$,
Eq.~(\ref{Psi-ei}), versus $g_{e}\rho_{e}$ at $g_{i}\rho_{i}$=0,
0.3, 0.6, and 0.8. Other parameters: $c=100$, $\Omega=20$.}
\label{fig2}
\end{figure}
%%%%%%%%%%%%%%%%%%%%%%%%%%%%%%%%%%%%%%%%%%%%%%%%%%%%%%%%%%%%%%%%%%%%%%
%%%%%%%%%%%%%%%%%%%%%%%%%%%%%%%%%%%%%%%%%%%%%%%%%%%%%%%%%%%%%%%%%%%%%%

The rate equations (\ref{ei}) look similar to the rate equations
derived in the pioneer works of Wilson and Cowan \cite{wc72,wc73}
who considered the dynamics of neural populations with excitatory
and inhibitory interactions. However, there are important
differences between our model and the Wilson-Cowan model. Our model
of interacting excitatory and inhibitory neurons is based on the
stochastic rules of activation and inactivation of individual
neurons (these are the rules (i)-(iv) in Sec.~\ref{model}) in
contrast to the deterministic phenomenological model in
\cite{wc72,wc73}. Using these rules, we derived the self-consistent
rate equations (\ref{ei}). Furthermore, Wilson and Cowan used as
relevant variables the fractions of excitatory and inhibitory
neurons which become active per unit time. Within our notations
these are $g_{e}\dot{\rho}_{e}$ and $g_{i}\dot{\rho}_{i}$,
respectively. In our approach in the case of classical random
graphs, the fractions of active excitatory and inhibitory neurons,
i.e., $g_{e}\rho_{e}$ and $g_{i}\rho_{i}$, are the relevant
variables.
%%It is important to note that choice of the relevant parameters to describe dynamics of neural networks is determined by the network
%%architecture. A correct choice permits us to derive a correct self-consistent equations. If the network structure differs from a classical
%%random graph then the relevant variables may be different from $\rho_{e}$ and $\rho_{i}$. One can prove this statement for neural networks on
%%the static model of scale-free complex networks \cite{gkk01}.
Also, on the base of experimental studies, Wilson and Cowan
postulated that the subpopulation response functions have a sigmoid
form. They used the standard mean field theory which neglects the
spatial heterogeneity, and assumed that all neurons are subjected to
the same average excitation of excitatory and inhibitory
populations. In our model, the functions
$\Psi_{e}(\rho_{e},\rho_{i})$ and $\Psi_{i}(\rho_{e},\rho_{i})$ in
Eqs.~(\ref{ei}) play the role of the response functions. We
calculated these functions exactly, taking into account the
heterogeneity of the classical random graph. According to
Eq.~(\ref{Psi-ei}), these functions have a sigmoid form with one
inflection point as a function of the parameter $g_{e}\rho_{e}$ in a
wide range of $g_{i}\rho_{i}$, see Fig.~\ref{fig2}. One can expect a
multimodal functional dependence with several inflection points if
there are several neural populations with different thresholds
$V_{th}$. Finally, in our stochastic approach, the set of
Eqs.~(\ref{s-ei}) permits the study of the dynamics of individual
neurons while Eqs.~(\ref{ei}) describe the global activity of the
neural populations. The Wilson-Cowan model only describes the global
activity of the neural populations.
%%However, it has been shown for many dynamical systems \cite{dgm08} that the standard mean field approximation may lead to wrong conclusions if
%%a network is strongly heterogeneous and has a heavy-tailed degree distribution similar to one found in real neural networks \cite{Sporns04}.
%%Our approach permits us to take into account exactly the heterogeneity.
Below we will show that the stochastic model as well as the
Wilson-Cowan model reveal hysteresis phenomena, decaying and
stable oscillations in neural activity.

\section{Steady states and avalanches}
\label{steady}
%%\emph{Steady states.}---

The steady states of the model are
determined by Eq.~(\ref{ei2}) at $\dot{\rho}_{a}=0$.
The steady solutions of Eq.~(\ref{ei2}) generalize the standard
bootstrap percolation
%%%%%%%%%%%%%%%%%%%%%%on undirected graphs
%%\cite{Chalupa79}
%%%%%%%%%%%%%%%%%
to a directed random graph with two types of vertices.
%%The standard
%%bootstrap percolation on a directed random graph corresponds to the
%%case $g_{e}=1$, $g_{i}=0$, and $Q_{e}=Q_{i}=0$ \cite{Breskin06,
%%Eckmann07, Soriano08,te09}.
%%
%%This state can found as follows. Let us chose at random
%%$F_{e}g_{e}N$ excitatory and $F_{i}g_{i}N$ inhibitory neurons on a
%%sparse random uncorrelated directed graph. They are activated and
%%fire forever. Then with probabilities $Q_{e}$ and $Q_{i}$ excitatory
%%and inhibitory neurons, including the neurons that  fire, are
%%removed at random from the network. Now the following update rule is
%%applied until the system reaches a steady state. A neuron becomes
%%active if the total input $V$, Eq.~(\ref{input}), from active
%%nearest neighbors is at least $\Omega$.
%%Notice that at $F_{e}=F_{i}=0$ the emergence of a non trivial
%%solution of Eq.~(\ref{steady}) means that activated neurons form
%%a giant $\Omega$-core in the considered directed network. This giant
%%subgraph of a graph is similar to the $k$-core of undirected complex
%%networks \cite{dgm06}. Recall that according to the standard
%%definition, the $k$-core of a graph is its maximum subgraph in which
%%each site  has at least $k$ neighbors \cite{Bollobas84}. $k$-cores
%%represent highly connected parts of a network.
%%Recently it was recognized that $k$-cores may play an important role
%%in dynamics of neural networks \cite{Schwab08}.
A particular case with $g_{i}=0$, $F_{e}=F_{i}$, and $Q_{e}=Q_{i}=0$
was studied in Refs.~\cite{Eckmann07}.
%%%%%%%%%%%%%%%%%%%%%%%%%%%%%%%%%%%%%%%%%%%%%%%%%%%%%%%%%%%%%%%%%%%%%
%%%%%%%%%%%%%%%%%%%%%%%%%%%%%%%%%%%%%%%%%%%%%%%%%%%%%%%%%%%%%%%%%%%%%%
\begin{figure}[t]
\begin{center}
\scalebox{0.25}{\includegraphics[angle=0]{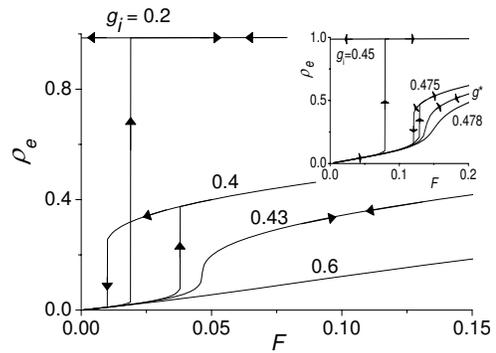}}
\end{center}
\caption{Activity $\rho_e$ of excitatory neurons versus the
activation parameter $F$ at different fractions of inhibitory
neurons $g_{i}$ from numerical solution of Eq.~(\ref{ei2}) at
$c=20$, $\Omega=3$. The jump and hysteresis disappear if $g_{i} >
g^{\ast} \simeq 0.43$. Arrows show increasing and decreasing $F$.
The insert shows results at $c=1000$, $\Omega=30$. Our simulations
confirm these results.} \label{fig3}
\end{figure}
%%%%%%%%%%%%%%%%%%%%%%%%%%%%%%%%%%%%%%%%%%%%%%%%%%%%%%%%%%%%%%%%%%%%%%
%%%%%%%%%%%%%%%%%%%%%%%%%%%%%%%%%%%%%%%%%%%%%%%%%%%%%%%%%%%%%%%%%%%%%%
Activation processes are shown in Fig.~\ref{fig3} at $F\equiv
F_{e}{=}F_{i}$, $Q_{e}{=}Q_{i}{=}0$ when $\rho_{e}=\rho_{i}$. One
can see that by increasing the activation parameter $F$, the
activity $\rho_{e}$ (and $\rho_{i}$) undergoes a jump at a critical
point $F_c$. A similar jump was observed in living neural networks
in vitro \cite{Eckmann07}.
%%which depends on the concentration $g_{e}$ of the excitatory neurons, the threshold $\Omega$, and the mean degree $c$.
If $F$ approaches $F_c$ from below, then
\begin{equation}
\rho_{a}=\rho_{a}^{(c)}-A(F_{c}{-}F)^{1/2}. \label{rc}
\end{equation}
where $A$ is a coefficient. This singular behavior evidences the
existence of long-range correlations between neurons and the
emergence of neural avalanches: the activation or deactivation of
one neuron triggers the activation or deactivation of a large
cluster of
%%excitatory and inhibitory
neurons. This phenomenon is similar to one that was found near the
point of emergence of a giant $k$-core \cite{dgm06}. Thus the
transition at $F_c$ is a hybrid phase transition (one which combines a jump
and a singularity). At $F=F_c$ the probability $G(s)$ that an
avalanche has a size $s$, including the activating neuron, is
\begin{equation}
G(s)\propto s^{-3/2}. \label{av}
\end{equation}
Similar neuronal avalanches were
%%predicted in [Eurich02] and then
observed in the cortex \cite{bp03,Plenz07}.
%% and in cortical networks developing in vitro \cite{pmbcm08}.
Using the approach from
\cite{dgm06}, we calculated $G(s)$ exactly at $g_{i}=0$ and $F\leq
F_{c}$:
\begin{equation}
G(s)=\frac{(n_{cr}s)^{s-1}}{s!}e^{-n_{cr}s}, \label{av2}
\end{equation}
where $n_{cr}$ is the average number of inactive subcritical
postsynaptic neurons of an inactive presynaptic excitatory neuron.
By definition, a subcritical neuron has exactly $\Omega-1$ active
presynaptic excitatory neurons. Successive activation of these
subcritical neurons forming finite clusters leads to avalanches. We
found that $n_{cr}=(1-F)d\Psi_{e}(\rho_{e},0)/d\rho_{e} \leq 1$
where $\rho_{e}$ is the neural activity in the steady state at a
given $F$. At the critical point $F=F_c$ we have $n_{cr}=1$. This
leads to Eq.~(\ref{av}) which we believe is also valid for
$g_{i}\neq 0$.

With increasing $g_i$ the size of the jump decreases. There is a
special critical point $g^{\ast}$ at which the jump is zero, and the
phase transition is continuous.
%%For example, we found that $g_s = 0.569$
%%%%$F_{c} =0.0464$
%%at $c=20$, $\Omega =3$.
%%At $g_{i} > g*$, there is no phase transition (see
%%Fig.~\ref{fig1}a).
There is no phase transition if $g_{i} > g^{\ast}$, or if $\Omega$
is larger than a critical threshold (see Fig.~\ref{fig3}).
%%In living neural networks one can influence
%%$\Omega$ and $g_i$ by chemicals \cite{Breskin06, Eckmann07,
%%Soriano08}.
%%These results are in qualitative agreement with experiments in
%%\cite{Breskin06, Soriano08} which revealed a jump of activity of
%%living neural networks at a critical value of electric stimulation.
In Fig.~\ref{fig3} we display numerical results for large mean
degree $c$=1000 and large $\Omega$=30, and for small mean degree
$c$=20 and small $\Omega$=3. Qualitatively the behavior is the same.
There is a range of $g_i$ in which the system demonstrates
bistability while the upper metastable state has activity $\rho_e$
not close to 1 (see small hysteresis loops in Fig. ~\ref{fig3}).
However, this region becomes smaller in the case of large $c$. This
indicates that with increasing $c$ and $\Omega$, this bistability
region decreases rapidly. Thus the hysteresis behavior crucially
depends on having finite values of $c$ and $\Omega$. In biological
systems the efficacy of inhibitory synapses is larger than that of
the excitatory ones. In our analysis we assumed that they are equal.
Our calculations show that an increase in magnitude of the
inhibitory efficacy moves the fraction $g_i$ of inhibitory neurons
at which the interesting bistability region takes place into a
region of biologically plausible values, namely about 0.2.

\section{Relaxation and oscillations}
\label{relaxation}

\begin{figure}[t]
\begin{center}
\scalebox{0.25}{\includegraphics[angle=0]{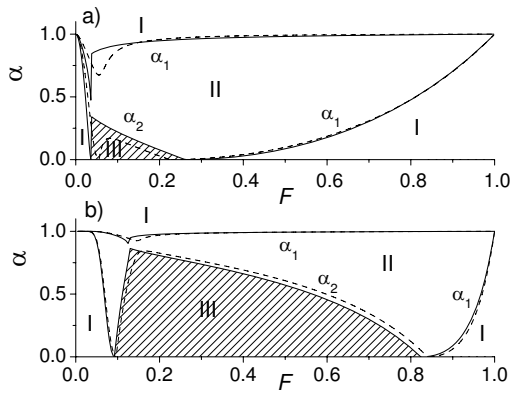}}
\end{center}
\caption{There are three regions on the $\alpha - F$ plane: (I) with
exponential relaxation; (II) with decaying oscillations; (III) with
stable oscillations.
%%Region (III) starts at $F=F_c$ if $g_{i} < g*$.
The boundaries $\alpha_{c1}$ and $\alpha_{c2}$, given by equations
Im$\gamma=0$ and Re$\gamma=0$, are shown at $g_{i}< g^{\ast}$ (solid
lines), and $g_{i} >  g^{\ast}$ (dashed lines).  (a) $c=20$,
$\Omega=3$, $g_{i}=$0.4 and 0.47. (b) $c=1000$, $\Omega=30$
$g_{i}=$0.475 and 0.478.} \label{fig4}
\end{figure}

Let us consider the relaxation of neural networks to a steady state.
We represent $\rho_{a}(t)$ as $\rho_{a}+ \delta \rho_{a}(t)$ where
$\delta \rho_{a}(t)/\rho_{a} \ll 1$, and $\rho_{a}$ is the
equilibrium activity of population $a$. Linearization of Eqs.
(\ref{ei2}) with respect to $\delta \rho_{a}(t)$ gives two coupled
linear equations:
\begin{equation}
\frac{d \delta \rho_{a}(t)}{\nu_{a}dt} {=} {-}\delta \rho_{a}(t){+}
D_{ae}\delta \rho_{e}(t){+}D_{ai}\delta \rho_{i}(t), \label{ei3}
\end{equation}
where $D_{ab}{\equiv}(1{-}F_a)(1{-}Q_a)\partial
\Psi_{a}(\rho_{e},\rho_{i})/\partial \rho_{b}$ for $a,b=e,i$. We
look for a solution in the form $\delta \rho_{a}(t)=A_{a} e^{-\gamma
t}$ with unknown $A_{a}$ and $\gamma$. The solution exists if the
determinant of this set of equations is zero. This condition gives
\begin{equation}
\gamma=\nu_{e}\{B_{1}{+}B_{2}{\pm}[(B_{1}{-}B_{2})^2{+}4\alpha
D_{ei}D_{ie} ]^{1/2} \}/2, \label{gamma}
\end{equation}
where $\alpha \equiv \nu_{i}/\nu_e$, $B_{1}=1{-}D_{ee}$,
$B_{2}=\alpha (1{-}D_{ii})$. Equation ($\ref{gamma}$) is valid in
the general case $\Psi_{e}\neq \Psi_{i}$. For the classical random
graph, using Eq.~(\ref{Psi-ei}), one can prove that $D_{ee},
D_{ie}>0$ while $D_{ei}, D_{ii} < 0$. Therefore $\gamma$ in
Eq.~(\ref{gamma}) may be a complex number in certain ranges of
parameters $c$, $g$, $F$, and $\alpha$. Where $ \texttt{Im}
\gamma=0$, relaxation is exponentially fast with the rate $\gamma$.
For example, at $\alpha=1$, we have $\gamma=\nu_{e}
(1{-}D_{ee}{-}D_{ii})\geq 0$. In this case $\gamma$ tends to 0 if
$F\rightarrow F_{c}$ from below as at a continuous phase transition.
However $\gamma$ is always finite above the critical point $F_{c}$.
If $\texttt{Re}\gamma>0$ and $\texttt{Im} \gamma \neq 0$, then
relaxation is in the form of decaying oscillations. If $ \texttt{Re}
\gamma <0$ and $\texttt{Im}\gamma \neq 0$, then any small deviation
from a steady state leads to oscillations around the state with an
increasing amplitude. However, in this case the linear
approximation, Eq.~(\ref{ei3}), is not valid, and it is necessary to
solve Eqs.~(\ref{ei2}). These three regions are shown in
Fig.~\ref{fig4}. We solved Eqs.~(\ref{ei2}) numerically in the case
$F_{e}=F_{i}=F$, $Q_{e}=Q_{i}=0$. We found that
%%at given $c$ and $\Omega$,
there is a region of $g_i$, which includes the special point
$g^{\ast}$, where $\texttt{Re}\gamma <0$ and $\texttt{Im}\gamma \neq
0$ if $0{<} \alpha {<} \alpha_{c2}{=}(D_{ee}{-}1)/(1{-}D_{ii}) <1$,
i.e., when inhibitory neurons have slower dynamics compared to the
dynamics of excitatory neurons.
%%Moreover, the activation parameter $F$ must be larger than $F_c$.
It turns out that in this region the neural system displays stable
oscillations around the steady state. Figure \ref{fig4} shows that
the larger the mean degree $c$ and the threshold $\Omega$ the
broader is the region with oscillations. We obtained similar
results for the model with synaptic delays. In particular, there is
a region of $g_i$ where oscillations emerge at $\alpha{=}1$
and $T_{ee}{=}T_{ei}{=}0$ if $T_{ie}{=}T_{ii}>T$ where $T$ is a
threshold. %%%%%%%%%%%%%%%%%%%%
The firing rate $\mu_{1}$ in human brains is typically in the range
1 - 400 Hz. In our model the frequency of oscillations $\omega_{o}$
is several times smaller than $\mu_{1}$. This gives $\omega_{o}$ in
the range of the waves observed in brain, i.e., $\omega_{o}
\lesssim$ 100 Hz.
%%%%%%%%%%%%%%%%%%%%%%%%%%%%%%%%%%%%%%%%%%%%%%%%%%%%%%%%%%%%%%%%%%%%%
%%%%%%%%%%%%%%%%%%%%%%%%%%%%%%%%%%%%%%%%%%%%%%%%%%%%%%%%%%%%%%%%%%%%%%
%%%%%%%%%%%%%%%%%%%%%%%%%%%%%%%%%%%%%%%%%%%%%%%%%%%%%%%%%%%%%%%%%%%%%%
%%%%%%%%%%%%%%%%%%%%%%%%%%%%%%%%%%%%%%%%%%%%%%%%%%%%%%%%%%%%%%%%%%%%%%

Replacing $f_{a}$ by $f_{a}(t){=}f_{a}{+}A_{a}\sin(\omega t)$ in
Eq.~(\ref{ei}), we study the response of the model,
$\rho_{a}{+}\Delta \rho_{a} \sin(\omega t{+}\varphi_{a})$, to a
small periodic stimulation, $A_{a} \ll f_{a}$. If $F$ approaches the
boundary between regions (II) and (III), see
%%curve $\alpha_{c2}$
Fig.~\ref{fig4}, the response
\begin{equation}
(\Delta \rho_{a}/A_{a})^2{\propto}
1/[(\omega {-}\texttt{Im} \gamma)^2{+}(\texttt{Re} \gamma)^2], \label{response}
\end{equation}
is enhanced because $\texttt{Re}\gamma{=}0$ at the boundary.
Therefore the transition from a state with incoherent neurons to a
state with global oscillations is a dynamical phase transition with
a sharp boundary (in the thermodynamic limit). In our model the
stochastic neural activity plays the role of noise while
interactions between neurons produce non-linear effects. Thus the
observed strong enhancement of the response is actually stochastic
resonance \cite{ghjm98,ma09}.
%%%%%%%%%%%%%%%%%%%%%%%%%%%%%%%%%%%%%%%%%%%%%%%%%%%%%%%%%%%%%%%%%%%%%
%%%%%%%%%%%%%%%%%%%%%%%%%%%%%%%%%%%%%%%%%%%%%%%%%%%%%%%%%%%%%%%%%%%%%%
\begin{figure}[t]
%%[tbhd]
\begin{center}
\scalebox{0.32}{\includegraphics[angle=0]{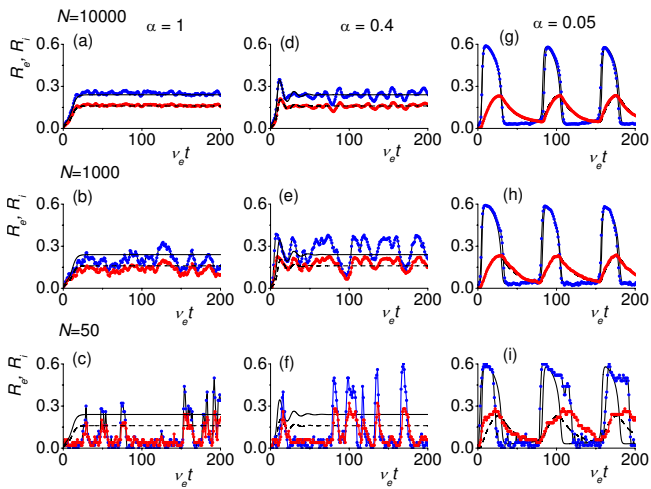}}
\end{center}
\caption{(color online). Fractions $R_{e}=\rho_{e} g_{e}$ and
$R_{i}=\rho_{i} g_{i}$ of active excitatory and inhibitory neurons
versus time. (a)-(c): $\alpha=1$ (region (I)). (d)-(f): $\alpha=0.4$
(region (II) ). (g)-(i): $\alpha=0.05$ (region (III)). Solid
(dashed) lines show theoretical $R_e$ ($R_i$) from Eqs.~(\ref{ei2}).
Blue (red) symbols refer to $R_e$ ($R_i$) from simulations at
$N=10000$ (1st row), 1000 (2nd row), and 50 (third row). %%We used
$F=0.05$, $g_{i}=0.4$, $c=20$, $\Omega=3$, $\mu_{2}=0$.} \label{fig5}
\end{figure}
%%%%%%%%%%%%%%%%%%%%%%%%%%%%%%%%%%%%%%%%%%%%%%%%%%%%%%%%%%%%%%%%%%%%%%
%%%%%%%%%%%%%%%%%%%%%%%%%%%%%%%%%%%%%%%%%%%%%%%%%%%%%%%%%%%%%%%%%%%%%%

\section{Simulations}
\label{simulations}

Our simulations supported the theoretical results. Random networks
with $N$ neurons were constructed by establishing directed links
between any neuron $i$ and neuron $j$, with probability $c/N$. In
the initial configuration all neurons were inactive. The state of
each neuron is then updated every $\Delta t$ time units (parallel
update) according to stochastic rules (i)-(iv). [Any other initial
configuration may be also used.]
%%After a short relaxation time the system reaches a steady state which do not depend on the initial state.]
The value of the time step $\Delta t$ was chosen such that the
probabilities $f \Delta t$, $\mu_{1} \Delta t$, and $\mu_{2} \Delta
t$ of the stochastic processes (i)-(iv) in Sec. \ref{model} for
excitatory and inhibitory neurons were sufficiently small. Reliable
results were obtained when these probabilities were about 0.1 or
smaller. Figure \ref{fig5} represents typical numerical results
obtained for systems of different sizes. All parameters used in
simulations are presented in the caption to Fig. 5. For a given
number of neurons $N$ we constructed several realizations of
networks, and then we simulated their stochastic dynamics, using the
rules (i)-(iv) in Sec.~\ref{model}. As one would expect, for the
considered stochastic model, different runs and different
realizations of neural networks differ slightly one from another.
With increasing $N$ these differences become smaller and smaller, so
these are standard run-to-run and realization-to-realization
variations.

Figure~\ref{fig5} shows a full set of regimes. One can see that in
regimes with exponential relaxation and decaying oscillations the
irregular activity of neurons decreases with increasing $N$. Already
at $N=1000$, a stimulation with $F > F_c$ activates a finite
fraction of neurons in agreement with the theory, though there are
strong irregular fluctuations around the steady state. In a small
network of 50 neurons stochastic effects are strong and suppress the
global activation. In Fig.~\ref{fig5} we also compare oscillations
predicted by Eq.~(\ref{ei2}) to our simulations. Interestingly,
these oscillations have a saw-tooth shape. Their period and shape
depend on the parameters of the model such as $F$, $\alpha$, $c$,
$\Omega$, and $g_i$. The theory and simulations are in very good
agreement at $N=10000$. Actually we found good agreement with only
$N=1000$. Surprisingly, the predicted oscillations emerge even in
small groups of 50 neurons where strong stochastic effects and non
negligible clustering could be expected. For $c=20$ and $N=50$ the
mean clustering coefficient is $C=c/N=0.4$ \cite{ab02,dgm08}, which
is close to the value $C=0.53$ found in the macaque visual cortex
\cite{Sporns04}. This intrinsic property of small groups of neurons
to oscillate may be very important for understanding communication
between neuronal groups in the brain \cite{Fries05}.

\section{Discussion}
\label{discussion}

First let us discuss the assumptions and approximations which are the basis of
our stochastic approach to noisy neural networks, and explain the biological
meaning of the model parameters from the point of view of
experimental and theoretical neuroscience.

In our model, activation of neurons by stimulus is a stochastic
process with a characteristic time which is equal to the reciprocal
rate $1/\mu_{1}$. In the brain, stochasticity in activation of
neurons by stimulus may appear in trial-to-trial variability of the
first-spike latency of neurons. The first-spike latency of a given
neuron is defined as the time from the onset of a stimulus to the
time of appearance of the first spike. The first-spike latency can
depend on many parameters. For example, for auditory neurons it
depends on the amplitude and frequency of stimulus \cite{heil04}. We
suppose that the reciprocal rate $1/\mu_{1}$ is of the order of the
mean first-spike latency of neurons. For simplicity, we assume that
$1/\mu_{1}$ is constant and does not depend on the input. The
first-spike latency may be of the order of the period of tonic
spiking or much larger if the input is near the threshold. In the
mammalian cortex the latency of regular spiking neurons for a
superthreshold input can be in the tens of milliseconds. This gives
$\mu_{1} \sim 10-400$ Hz.

In our approach it is assumed that each neuron may be active
spontaneously, that is, it may discharge without
experimenter-controlled stimulations. At the present time, the
mechanisms and functional significance of spontaneous neural
activity are not well understood and it is a topical problem in
experimental and theoretical neuroscience
\cite{Ermentrout08,Faisal08,AmitBrunel997,dc05}. The typical
spontaneous background activity observed in the cortex is 1-5
spikes/s. Interactions between neurons play an important role in
this activity \cite{AmitBrunel997}. Spontaneous activity in the
brain may be mediated by intrinsic, intracellular, generated
activity and circuit feedback mechanisms. Neural activity in one
region of the brain may propagate to other regions, circulating in
recurrent loops. For example, neurons in the thalamus and the
cerebral cortex form recurrent loops \cite{steriade00}. The study of
spontaneous activity in neocortical slices \cite{mao01} gives
evidence that supports both mechanisms. For our model a mechanism of
spontaneous activity is unimportant. One can assume that spontaneous
activity takes place by the intrinsic mechanism. Alternatively one
can consider the neural network as part of a large system from which
neurons receive random inputs. In real neural networks only a
fraction of neurons are spontaneously active \cite{mao01}. In the
present paper we study the case in which all excitatory and
inhibitory neurons may be spontaneously active. One can show that if
only some fraction of neurons is spontaneously active, the dynamics
of the neural networks would be qualitatively the same.
%%In this system a significant percentage of spontaneously active neurons are mostly the pyramidal cells \cite{mao01}.

Furthermore, we considered the activation of neurons by an external
stimulus as a stochastic process. Experimental work supports this
assumption. For example, it was revealed that a moving whisker can
have only a 15\% chance of generating spikes in a neuron in the
mouse somatosensory cortex \cite{bs02}. Unfortunately, much less is
known about stochastic processes of spontaneous deactivation of
neurons. In our model, the reciprocal rate $1/\mu_{2}$ is the
characteristic time at which neurons stop firing due to irregular
fluctuations on the input or due to random processes taking place
inside the cells. Recently it was shown that spontaneous activity of
single neurons may be driven by noise which can not only activate
but can also inhibit spiking activity of neurons of both type 1 and
type 2 \cite{gjt08,tgj09,tj09}.

The rates of the stochastic processes discussed above can be found
from statistical analysis of activation and inactivation events in
neural networks. They can also be measured by use of the patch-clamp
technique: one can stimulate presynaptic excitatory and inhibitory
neurons and then measure the probability of activation of a
postsynaptic neuron through the distribution of first-spike timing times.
%%By this method one can also measure distribution of
%%times at which the postsynaptic neuron stops firing spontaneously.
%%From these data one can estimate $\mu_{2}$.

The proposed stochastic model is not restricted to regular spiking
neurons. One can also analytically study noisy neural networks with
neurons which generate random spike trains, for example, Poisson
spike trains as found in recordings from neurons \emph{in vivo} and
\emph{in vitro} \cite{Faisal08}. The proposed stochastic approach
can also be generalized to study analytically neural networks with
neurons having the type 1 and 2 dynamical behavior shown in
Fig.~{\ref{fig1}} for the case when correlations between presynaptic
neurons may be neglected. However these generalizations are out of
the scope of the present paper.

We found that even a small group of neurons reveals intrinsic
oscillations which are robust against strong stochastic fluctuations
(see Fig.~\ref{fig5}). It means that despite noise, neurons in a
small group can synchronize their dynamics. We believe that this
result opens interesting possibilities to study and model
communication between different groups of neurons and the
transmission of activity from one group of neurons to another. In
the recent review \cite{Fries05}, ``neuronal communication between
neuronal groups through neuronal coherence'' was considered as a
mechanism for cognitive dynamics. On the basis of neurophysiological
data, Fries suggested that coherently oscillating neuronal groups
can interact effectively \cite{Fries05}. This idea is based on an
assumption that activated neuronal groups have an intrinsic tendency
to oscillate. Our model supports this assumption, showing that
oscillations indeed are an intrinsic property
%%of small neural groups
and robust against noise. In our model one can model communication
between neural populations or neural modules. Synchronization of
neurons or groups or modules of neurons in the regime with
oscillations can play an important role in this communication.
Indeed, our preliminary simulation of interacting neural communities
reveals complex patterns of neural activities. On the basis of our
stochastic model one could study the computational role of network
oscillations and how oscillations contribute to the representation
of information \cite{sp06}.

Real neural networks have a scale-free degree distribution
\cite{Sporns04} rather than a simple Poisson distribution. A
preliminary study of neural networks with a scale-free degree
distribution showed that these networks demonstrate dynamical
properties qualitatively similar to properties of the networks
studied above.

Let us discuss possible experiments to test the proposed model.
First, it would be interesting to observe hysteresis and neural
avalanches like those found in Sec.~\ref{steady} near the
discontinuous phase transition. A similar discontinuous phase
transition was revealed in activations of living neural networks by
a stimulus in recent works \cite{Eckmann07}. Neural avalanches in
these biological systems can be found by use of microelectrode
arrays as described in \cite{bp03,Plenz07}. Second, the theory
predicts that the emergence of global oscillations is a dynamical
phase transition. A strong enhancement of the response of a neural
network to a periodic stimulus in the range of frequencies of these
oscillations manifests this transition. These oscillations can be
driven by an external stimulus, neural pacemakers or noise. Though
we demonstrated this behavior for ideal neurons,
%%of type 2 excitability,
we believe that it is a universal critical phenomenon if a
sufficiently large group of neurons is involved in these
oscillations. It would be interesting to observe experimentally this
enhancement which in fact is stochastic resonance. This enhancement
may be found, for example, in experiments similar to the experiments
carried out by Fries \emph{et al.} \cite{frrd01} who observed that
neurons of macaque monkeys activated by the attendant stimulus show
increased gamma-frequency (35 - 90 Hz) oscillations. One can expect
that a response of the neurons to periodic stimulus in the
gamma-band frequency will be enhanced near critical attention above
which gamma-frequency oscillations emerge.

\section{Conclusion}
\label{conclusion}

In conclusion, based on experiments and ideas of cellular automata
we developed a model of noisy neural networks with excitatory and
inhibitory neurons and a complex network architecture. We considered
neurons which are either inactive or fire a regular train of spikes
with a given frequency (neurons with type 2 dynamical behavior). In
this model we took into account spontaneous neural activity, which
plays the role of noise, the activation of neurons by a stimulus,
neural pacemakers, and interactions between neurons. We derived rate
equations describing the evolution of the global neuronal activity.
These equations are exact for infinite uncorrelated complex networks
with arbitrary degree distributions, though for brevity, we
presented results only for classical random graphs.
%%though we found qualitatively similar results for a scale free networks.
This model has a complex phase diagram with self-organized active
neural states, hybrid phase transitions, hysteresis phenomena and a
rich array of behavior including decaying and stable oscillations,
stochastic resonance, and neural avalanches. We showed that global
oscillations and stochastic resonance are intrinsic properties of
this non-linear dynamical system. The oscillations emerge when
noise, i.e., the spontaneous neural activity, reaches a threshold
level while stochastic resonance is a precursor of global
oscillations. We also found that the network structure is important.
The larger the connectivity the broader is the region with global
oscillations. Our simulations revealed that even small groups of
50-1000 neurons display oscillations similar to large networks.
%%Our simulations revealed that oscillations are %%robust against stochastic fluctuations and
%%are an intrinsic property of even small groups of neurons.
%%50-1000 neurons display oscillations similar to large networks.
%%despite strong stochastic fluctuations.
%%which usually are strong in small networks.
%%This demonstrates the robustness of oscillations of small
%%groups of neurons against noise.
%%The proposed model can also have other applications, for example, rumor spreading in social networks with two
%%types of agents: susceptible (excitatory) agents and sceptics (the
%%inhibitors).
%%We believe that the simplicity of this model will facilitates
%%investigations of the role of the network structure in dynamics of
%%neural networks.
%%The proposed model explains qualitatively recent experimental
%%investigations \cite{Eckmann07} which revealed a first order phase
%%transition in an activation of living neural networks by a stimulus.
%%Results are represented in Fig. 1 and in the text on page 3.
%%This model reveals neural avalanches which are similar to neuronal
%%avalanches observed in the cortex \cite{Plenz07}, see Eq. (10).

Further development of the model can be done by taking into account
the real structure of neural networks (clustering, degree-degree
correlations, modular structure, and other structural properties), a
dependence of firing rate on input, variability of synapses,
evolution of network structure, for example, considering growing
networks, or variable strength of synapses, and so on. Apart from the
perspectives discussed above, one can also apply this stochastic model to
study communication between different groups of neurons and the
transmission of activity from one group or module of neurons to
another, taking into account noise and complex network architecture.

\begin{acknowledgments}

This work was partially supported by projects POCI: SAU-NEU/103904,
BIA-BCM/62662, FIS/71551, FIS/108476,
%%PTDC/FIS/71551/2006, PTDC/SAU-NEU/103904/2008
and the ARTEMIS and SOCIALNETS EU projects. The authors thank D.
Holstein for help in simulations and G. Baxter for help in preparing
the manuscript.

\end{acknowledgments}

%%\newpage

\end{document}